\documentclass[reprint,superscriptaddress,amsmath,amssymb,prl,nolongbibliography]{revtex4-2}
\usepackage{graphicx}
\usepackage{xcolor}
\usepackage{dcolumn}
\usepackage{bm}
\usepackage{soul}
\newcommand{\wpp}{\omega_{\rm p}}
\newcommand{\dd}{{\rm d}}

\begin{document}

\preprint{APS/123-QED}

%%%%%%%%%%%%%%
% Title
%%%%%%%%%%%%%%
\title{Fast radio bursts as precursor radio emission from monster shocks}%

%%%%%%%%%%%%%%
% Authors
%%%%%%%%%%%%%%
\author{A. Vanthieghem}
\email{arno.vanthieghem@obspm.fr}
\affiliation{Sorbonne Universit\'e, Observatoire de Paris, Universit\'e PSL, CNRS, LERMA, F-75005, Paris, France}
\author{A. Levinson}
\affiliation{School of Physics and Astronomy, Tel Aviv University, Tel Aviv 69978, Israel}

%%%%%%%%%%%%%%
% Date
%%%%%%%%%%%%%%
\date{\today}

%%%%%%%%%%%%%%
% Abstract
%%%%%%%%%%%%%%
\begin{abstract}
It has been proposed recently that the breaking of MHD waves in the inner
magnetosphere of strongly magnetized neutron stars can power different types of
high-energy transients. Motivated by these considerations,
we study the steepening and dissipation of a strongly magnetized fast magnetosonic wave 
propagating in a declining background magnetic field, by means of 
particle-in-cell simulations that encompass MHD scales.   
Our analysis confirms the formation of a monster shock as $B^2-E^2 \to 0$, that dissipates
about half of the fast magnetosonic wave energy. 
It also reveals, for the first time,  the generation of a high-frequency precursor wave by %a synchrotron maser instability at 
the monster shock, carrying a fraction of $\sim 10^{-3}$ of the total energy dissipated at the shock. 
The spectrum of the precursor wave exhibits several sharp harmonic peaks, with frequencies in the GHz band 
under conditions anticipated in magnetars. Such signals may appear as fast radio bursts.

\end{abstract}

\maketitle

%%%%%%%%%%%%%%
% Introduction
%%%%%%%%%%%%%%
The propagation and dissipation of large amplitude waves in strongly magnetized plasma is an issue of considerable interest in 
high-energy astrophysics.  Such waves have long been suspected to be responsible for immense cosmic eruptions, including
magnetar flares \cite{thompson1995,parfrey2013,Chen22,yuan22,beloborodov23,mahlmann23}, fast radio bursts (FRBs) \cite{lyubarsky14,lyutikov2016,waxman2017,lyubarsky2021,mahlmann22,thompson23}, delayed gamma-ray 
emission from a collapsing magnetar \cite{most2024monster},
X-ray precursors in BNS mergers \cite{beloborodov23}, 
and conceivably gamma-ray flares from blazars and other sources. 

Disturbance of a neutron star magnetosphere, e.g., by star quakes, collapse or collision with a compact companion, generates MHD waves that propagate in the magnetosphere. 
In general, both Alfv\'en and magnetosonic modes are expected to be produced during abrupt magnetospheric perturbations, with millisecond periods - a fraction of the stellar radius \cite{beloborodov23}.  
The amplitude of such a wave is usually small near the stellar surface, but gradually grows as the wave propagates down 
the decreasing background dipole field.  
As the wave enters the nonlinear regime, it is strongly distorted.  A periodic fast magnetosonic (FMS) wave, in particular, 
steepens and eventually forms a shock (termed monster shock) when the wave fields reach a 
value at which $B^2 - E^2 \approx 0$ \cite{Chen22,beloborodov23}.  Half of the energy 
carried by the wave is dissipated in the shock and radiated away, producing a bright X-ray burst. 
The other half can escape the inner magnetosphere without being significantly distorted.  This is also true in the case of
FMS pulses in which the electric field does not reverse sign. 
The escaping wave (or pulse) can generate a fast radio burst (FRB), either by compressing the magnetar current sheet \cite{lyubarsky2020,mahlmann23}, or 
through a maser shock produced far out via collision of the FMS pulse with surrounding matter~\cite{lyubarsky14,waxman2017,metzger2019,lyubarsky2021,khangulyan2022}.  As will be shown below, GHz waves 
are also produced at the precursor of monster shocks and may provide another production mechanism for FRBs.

Steepening and breakdown of FMS waves may also be a viable production mechanism of rapid gamma-ray flares in blazars.
In this picture, episodic magnetic reconnection in the inner magnetosphere, 
at the base of magnetically dominated jet \cite[e.g.,][]{ripperda2022}, can 
excite large amplitude MHD waves that will propagate along the jet, forming radiative shocks close to the source,
that can, conceivably, give rise to large amplitude, short duration flares.  
Whether detailed calculations support this picture remains to be investigated. But if this mechanism operates effectively in black hole jets, it can alleviate the issue of dissipation of relativistic force-free jets.  
Striped jets can also produce rapid flares, but the mean power of such jets is likely to be considerably smaller than that of jets produced
by ordered fields in MAD states \cite{mahlmann20,chashkina21}.  Dissipation of ordered fields must rely on instabilities, like the current driven kink instability, which 
are expected to be too slow to account for the rapid variability seen in many blazars, if generated at all. 

In this Letter, we study the steepening and breaking of a FMS wave by means of first principles plasma simulations.
The novelty of our study is the disclosure of a high-frequency precursor wave, generated by the monster shock, that
we propose might explain some of the enigmatic fast radio bursts and perhaps other radio transients.
Previous simulations \cite{Chen22} already demonstrated the steepening of a nonlinear FMS wave, however, due to the
small scale separation used, it was not possible to properly follow the wave evolution and resolve the shock.  In contrast, 
our simulations can follow the wave evolution in the MHD regime and  resolve the shock structure, owing to the large scale 
separation applied.
%By comparing simulation results with an analytic model, we are able to derive scaling relations of fundamental properties. 
%As hinted above, we also identify, for the first time, the generation of a high-frequency precursor wave by a synchrotron maser instability at the monster shock. 
%We propose that it might explain some of the enigmatic fast radio bursts and perhaps other radio transients.

\begin{figure*}
  \centering
  \includegraphics[width=1\textwidth]{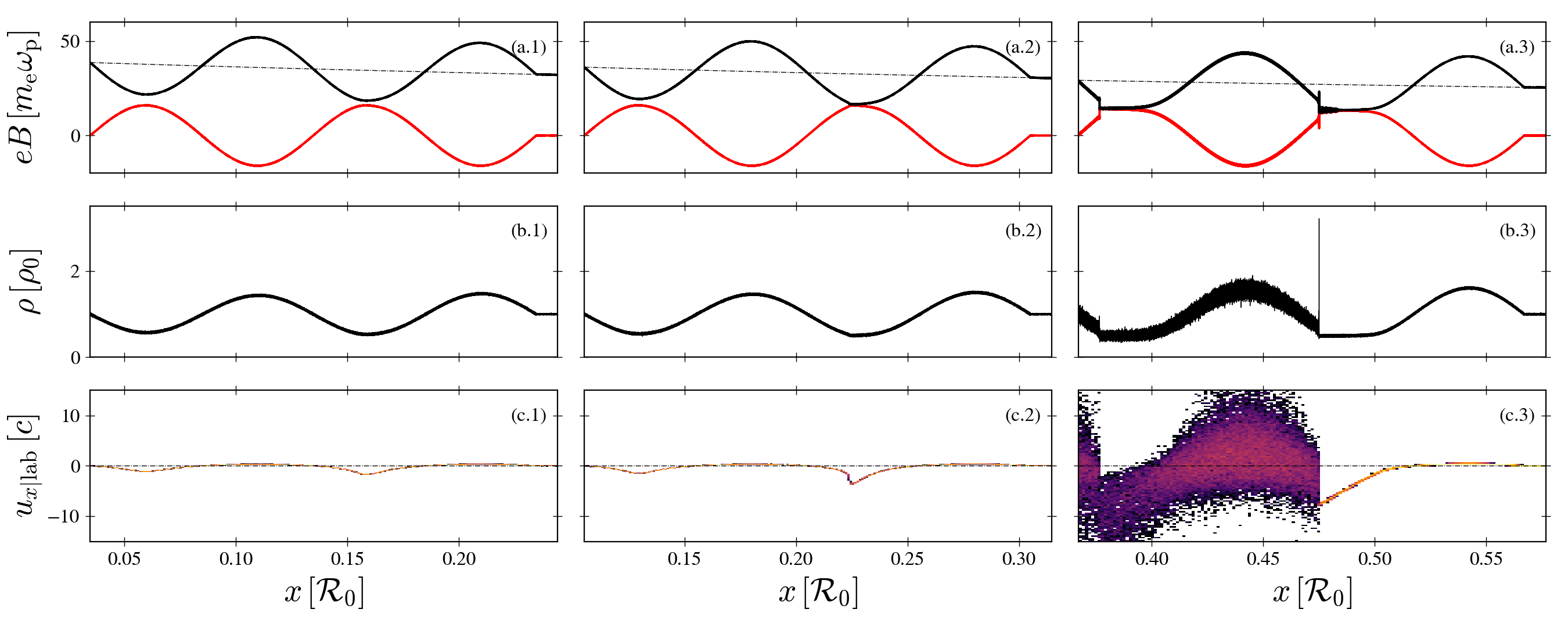}
  \caption{Three snapshots from the evolution of the FMS wave, taken before wave breaking at $t\,=\, 0.23\,\mathcal{R}_0/c$ (left), during shock formation at $t\,=\, 0.30\,\mathcal{R}_0/c$ (middle), and well after shock formation at $t\,=\, 0.57\,\mathcal{R}_0/c$ (right). Panels (a) display $B_z$ (black line), $-E_y$ (red line), and $B_{\rm bg}$ (black dot-dashed line), panels (b) show the density, and the lower panels (c) show the distribution of the longitudinal plasma 4-velocity in the laboratory frame $u_{x|\rm Lab}$, with the black dot-dashed line indicating the flow velocity upstream of the wave. Shock formation is first observed at $x\,=\,0.224\,\mathcal{R}_0$ in panel (c.2). A second shock subsequently forms and is seen at $x\,=\,0.377\,\mathcal{R}_0$ in panel (c.3).}
  \label{fig:steepening}
\end{figure*}

%%%%%%%%%%%%%%
% Main
%%%%%%%%%%%%%%
We begin by analyzing the general properties of nonlinear wave propagation and the onset of wave breaking \footnote{It is worth noting that our
derivation of the characteristics equations, detailed in the appendix, is general and not restricted to the  regime of $\sigma_{\rm bg}\,\gg\,1$, as in \citep{beloborodov23}.}. 
We consider a FMS wave propagating in a medium having a background magnetic field $\pmb{B}_{\rm bg} = B_{\rm bg} \hat{z}$,
and proper density $\rho_{\rm bg}$, where $B_{\rm bg}$ and $\rho_{\rm bg}$ may depend on $x$.  The magnetization of the background medium is defined as 
\begin{equation}\label{eq:sigma_bg}
\sigma_{\rm bg} = \frac{B^{2}_{\rm bg}}{4\pi \rho_{\rm bg}},
\end{equation}
with $\sigma_{\rm bg} \gg 1$ for the force-free magnetospheres under consideration here. 
The wave is injected at time $t=0$ from some source located at $\pmb{x}_0$, 
and propagates in the positive $x$ direction.  The electric and magnetic field components are given, respectively, by
$\pmb{E} = E(x,t)\hat{y}$ and $\pmb{B}_{\rm w} = \pmb{B} - \pmb{B}_{\rm bg} = B_{\rm w}(x,t) \hat{z}$.
For radial propagation in the equatorial plane of a dipole field $(\hat{x},\hat{y},\hat{z})=(\hat{r},\hat{\phi},\hat{\theta})$.
In the limit of ideal MHD, $F_{\mu\nu}u^\nu$=0, where $F_{\mu\nu}$ is the electromagnetic tensor and $u^\mu = (\gamma, \gamma v, 0,0)$ the plasma 4-velocity,
the plasma 3-velocity equals the drift velocity: $\pmb{v} = \pmb{E}\times \pmb{B}/B^2 = E/B \, \hat{x}$. The corresponding Lorentz factor is $\gamma = B/\sqrt{B^2-E^2}$.
The simple wave considered here moves along the characteristic $C_+$, defined by:
\begin{equation}\label{eq:dx+_a}
\frac{dx_+}{dt} \equiv v_{+} =\frac{v + a}{1 + a v},
\end{equation}
where $a$ is the fast magnetosonic speed measured in the fluid rest frame.  In the appendix, it is shown that for a cold plasma 
$a(\lambda) = \tanh (\lambda/2+c)$ and $v_{+}(\lambda) = \tanh(3\lambda/2+ c)$, where $c = \ln \left(\sqrt{\sigma_{\rm bg}} +\sqrt{\sigma_{\rm bg}+1}\right)$
and
\begin{equation}
\lambda\equiv \frac{1}{2 }\ln\left(\frac{1+v}{1-v}\right).
\end{equation}
For a uniform background field, ${\rm d}B_{\rm bg}/{\rm d}x=0$, $\lambda$ is conserved along the characteristic $C_+$ (a Riemann invariant).
However, in general, $\lambda$ varies along $C_+$ with a dependence on the profile of $B_{\rm bg}$.

For a cold plasma, the wave magnetization relates to $a$ through $\sigma= a^2/(1-a^2) = \sinh^2(\lambda/2+c)$.  In the regime $\sigma_{\rm bg}\gg 1$ 
this approximates to:
\begin{equation}\label{eq:sig_lambda}
\sigma(\lambda) \approx \sigma_{\rm bg}(\lambda) e^\lambda = \sigma_{\rm bg}(\lambda) \sqrt{\frac{1+v}{1-v}}.
\end{equation}
 Since for ideal MHD, the continuity equation 
implies that $B'/\rho$ is conserved, with $B'= B/\gamma$ being the magnetic field measured in the fluid rest frame, one finds a compression factor 
of $B'/B_{\rm bg} = \rho/\rho_{\rm bg}  = e^\lambda$, and $B/B_{\rm bg} = \gamma B'/B_{\rm bg} = (1-v)^{-1}$.  The relations $E/B= v$ and $B_w = B-B_{\rm bg}$ then yield:
\begin{equation}\label{eq:wave_fields_sec}
E = B_{\rm w} = \frac{v(\lambda)}{1-v(\lambda)} B_{\rm bg}.
\end{equation}
Note that $E$ can vary between $E=-B_{\rm bg}/2$ at $v=-1$ ($\lambda \to -\infty$) and $E \to \infty$ at $v\to 1$ ($\lambda\to\infty$).
Note also that $(B^2-E^2)/B_{\rm bg}^2 = e^{2\lambda}$. Thus, $B^2-E^2 \to 0$ as $\lambda\to -\infty$ or $E\to -B_{\rm bg}/2$.

In cases where the background field declines along the characteristic $C_+$, viz., ${\rm d}B_{\rm bg}/{\rm d}x<0$, energy conservation implies that 
$|E|/|B_{\rm bg}|$ increases.  For instance, for a wave propagating in the inner magnetosphere of a neutron star, which
to a good approximation is a dipole, $|E|/|B_{\rm bg}| \propto r^2$.  This means that $\lambda$ and, hence, 
$B^2-E^2$, decrease along $C_+$, ultimately approaching $B^2-E^2 = 0$. 
From the expression for the wave velocity given below Eq. \eqref{eq:dx+_a}, it is seen that $v_+ =0$ at $\lambda = -2c/3$, or equivalently $(B^2-E^2)/B_{\rm bg}^2 = (4\sigma_{\rm bg})^{-2/3}$,
$\gamma v = \sinh \lambda \approx - (4\sigma_{\rm bg})^{1/3}$.
At even smaller $\lambda$ values $v_+$ changes sign and the characteristic turns back towards the injection point.  Consequently, wave
breaking is anticipated around this location.  The exact value of $\lambda$ at the moment of shock birth, $\lambda_{\rm s}$, depends on the details.  
For example, in case of a periodic planar wave with electric field $E(x_0,t)= E_0\sin(\omega t)$ at the boundary $x_0$, 
propagating in a background magnetic field $B_{\rm bg} =B_0 (x_0/x)$, we
find $\lambda_{\rm s} = \ln(\omega^2 x_0^2/16 a \sigma_{\rm bg}^2)^{1/4}$, where $a=E_0/B_0$ - see appendix for details.
As the wave propagates,  the plasma upstream of the shock continues to accelerate backward ($\gamma v = \sinh \lambda <0$), 
the magnetization declines (Eq. \ref{eq:sig_lambda}), and the shock strengthens.  Only wave phases that satisfy $E > -B_{\rm bg}/2$ survive. The other parts are erased through shock dissipation.

We illustrate the analytical model and study the kinetic evolution of the wave after shock formation with 
1D PIC simulations performed with the relativistic electromagnetic code Tristan-MP V2~\cite{Tristan_v2}.
At $\wpp t\,=\,0$, we launch a periodic FMS wave from $x=0$, letting it propagate along $+\hat{x}$ in a cold pair plasma with a declining static and external background magnetic field, $\pmb{B}_{\rm bg}(x) = B_0\,(1+x/\mathcal{R}_0)^{-1}\,\hat{z}$, constant 
density and initial background magnetization $\sigma_{0}= 1600$, so that  $\sigma_{\rm bg}(x) = 
\sigma_0\,(1+x/\mathcal{R}_0)^{-2}$. To capture MHD scales, the ratio of the wavelength, $\lambda_w$, to skin depth, $c/\wpp=\sqrt{m_{\rm e} c^2/4 \pi n_{\rm bg} e^2}$, was
taken to be $\lambda_w\,\wpp/c\,=\,2\,\wpp/\omega\,=\,1.06\times10^4$ with 60 cells per skin depth, $c\Delta t\,=\,\tfrac{1}{2}\Delta x$, and 50 particles per cell. The amplitude of the wave is set to $0.4\,B_0$.
The gradient length scale of the background magnetic field, $\mathcal{R}_0$, is $\mathcal{R}_0\,=\,10\lambda_w\,\simeq\,1.06\times10^5\,c/\wpp$. 
For further details on the derivation and numerics, see the appendix.

Figure~\ref{fig:steepening} illustrates the propagation in the laminar (left), steepening (middle), and shock wave (right) regimes along the gradient. We observe the onset of strong steepening around
$x\,=\, 0.211\,\mathcal{R}_0$. In the steepening region where $B_z\gtrsim-E_y$ [panel (a.2)], the flow accelerates up to relativistic speeds [panel (c.2)]. The wave then breaks, forming a shock, at 
$x_{\rm s}\,=\,0.225\,\mathcal{R}_0$, corresponding to about 2.25 wavelengths from the injection point. A second shock subsequently forms [Fig.~\ref{fig:steepening}(a-b.3)].
The shock forms near the wave trough, as expected from the analytic derivation in the appendix. Panels~\ref{fig:steepening}(a-b.3)  also reveal a soliton structure, further detailed below, characterized by a sharp peak in density and electromagnetic field at the shock and associated with significant bulk flow heating.
A careful examination indicates that the value of the compression factor
at wave breaking is $\lambda_{\rm s} = \ln \, (0.1)$, corresponding to a plasma 4-velocity of $u_x = -5$.   The corresponding value
 of the invariant at this location is $(B^2-E^2)/B_{\rm bg}^2 = 10^{-2}$. These values are in good agreement with 
the analytic results derived in the appendix, but note that the density profile adopted there differs from the one employed in the simulations.
Following shock formation, the wave develops a plateau at phases where $E = -B_{\rm bg}/2$. This is clearly 
seen in Fig.~\ref{fig:steepening}(a.3).  This plateau extends in size over time until complete eradication of the lower part of the wave. 
The corresponding wave energy is dissipated at the shock. The plasma in
the plateau accelerates towards the shock before crossing it, while the magnetization decreases [see Fig.~\ref{fig:steepening}(c.3)]. The Lorentz factor just upstream of 
the shock increases as the wave evolves, reaching a maximum towards the end of the simulations, and then starts declining.

\begin{figure}
  \centering
  \includegraphics[width=1\columnwidth]{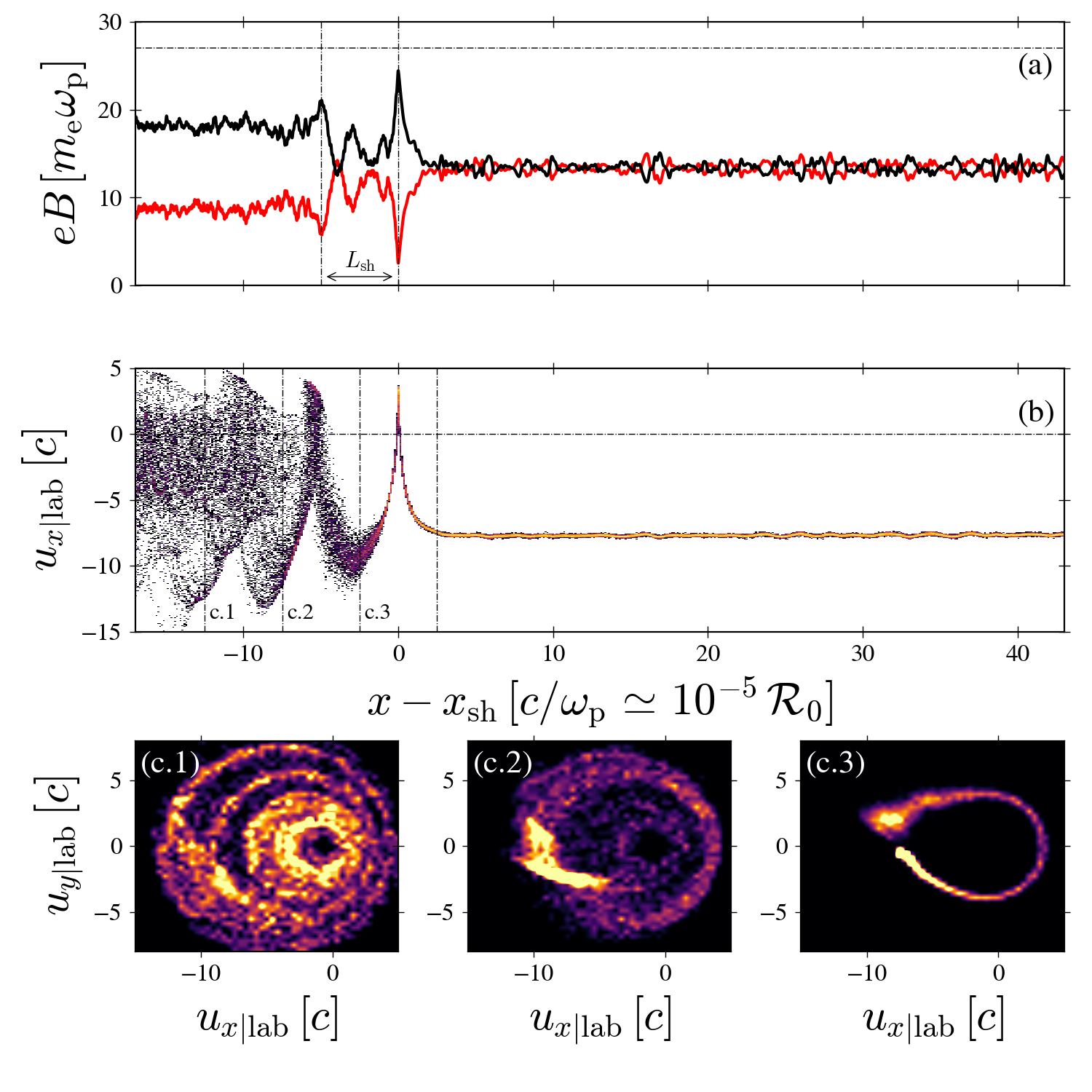}
  \caption{Close up on the shock structure in the steepening zone. The shock is centered on the leading soliton-like structure. 
  (a) Profile of $B_z$ (black line), $-E_y$ (red line), $B_{\rm bg}$ black dot-dashed line) . The black dot-dashed line shows the background magnetic field. Vertical lines indicate the positions of the two leading solitons.
  (b) Distribution of the longitudinal 4-velocity along the shock profile. The vertical lines delineate the corresponding space over which the phase-space profiles of lower panels (c.1-3) are taken.
  (c) Phase-space profile of the electrons. 
  The total $u_x-u_y$ electron distributions corresponding to the three subsections are respectively 
  displayed on insets (c.1), (c.2), and (c.3).}
  \label{fig:shock}
\end{figure}

\begin{figure}
  \centering
  \includegraphics[width=1\columnwidth]{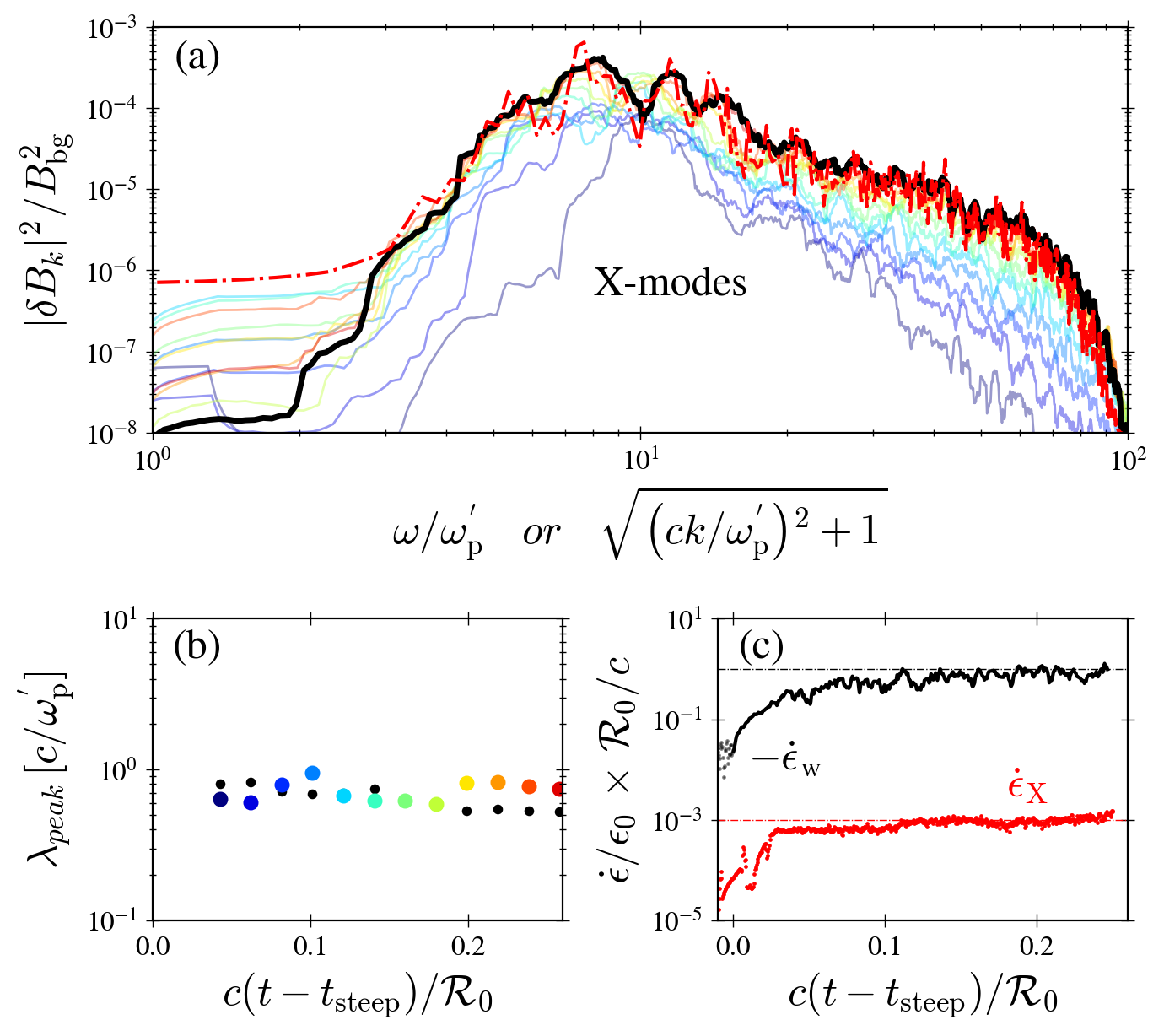}
   \caption{
  (a) k-spectra of the X-mode (colored solid lines), and $\omega$-spectrum of the X-mode (red dot-dashed line), computed in the source frame just upstream of the shock. The $k$-space spectra cover the domain $\left( x - x_{\rm sh} \right)\,=\, \left[ 10,\,210 \right]\,c/\omega_{\rm p}$ in the time interval $c (t-t_{\rm steep})/\mathcal{R}_0\,=\,[0.04,\,0.28]$. The colors of the solid k-spectra lines correspond to the time of measurement, as indicated in panel (b). The thick, solid black line delineates the converged k-spectrum, and the red dot-dashed line is the space-averaged $\omega$-spectrum. (b) Comparison between the highest peak wavelength (colored circles) and $L_{\rm sh}/3$ (small black circles),  at different times, where $L_{\rm sh}$ is the approximate distance between the two leading solitons (Fig. \ref{fig:shock} top panel). (c) Black: Total averaged dissipation rate $\dot{\epsilon}_{\rm w}$ of the FMS wave energy at the shock averaged over one wavelength and normalized by $\epsilon_0 c /\mathcal{R}_0$, where 
 $\epsilon_0$ is the initial FMS wave energy stored over half a wavelength;
 Red: Energy pumping rate into precursor wave dominated by X-modes, averaging around $\dot{\epsilon}_{\rm X}/\epsilon_0\,=\,10^{-3}\,c/\mathcal{R}_0$.}
  \label{fig:spectrum}
\end{figure}

Figure \ref{fig:steepening} also reveals the generation of high-frequency precursor waves, %likely formed by a synchrotron maser 
identified earlier in planar shocks under somewhat different conditions \cite{sazanov73,Hoshino_1992,Gallant_1992,lyubarsky2006,iwamoto2017,iwamoto2018,plotnikov2019}.
Its emission mechanism is yet unresolved \footnote{In previous publications, the generation of the precursor wave
has been attributed to the synchrotron maser instability. To us, the fact that the spectrum of this wave exhibits sharp harmonics associated
with the distance between the solitons (Fig. \ref{fig:spectrum}) suggests that the emission mechanism may actually be different.}.
%This wave is seen only in the leading shock, at $x\,=\,0.475\,\mathcal{R}_0$ in Fig.~\ref{fig:steepening}(c.1-3).
This wave is seen only as high-frequency modulation in the leading shock, at $x\,\gtrsim\,0.475\,\mathcal{R}_0$ in Fig.~\ref{fig:steepening}(a.3).
The reason why the trailing shock does not generate a precursor wave is the high temperature of the upstream plasma caused by the 
passage of the leading shock (see panel c.3 in Fig. \ref{fig:steepening}), consistent with earlier findings  \cite{babu2020}.
%the wave emission mechanism operates efficiently  only when the thermal spread of emitting particles is sufficiently small \cite{babu2020}.  
We emphasize that under realistic conditions, 
fast cooling of the shocked plasma, which can change the conditions at the trailing shock, is anticipated \cite{beloborodov23} 
but ignored in the present analysis. We further note that even if the background  plasma is preheated, 
the strong decompression of the accelerated plasma ahead of the leading shock will likely lead to rapid adiabatic cooling that will enable
generation of the precursor wave.
To elucidate the basic features of the shock structure and the precursor wave, we present in Fig. \ref{fig:shock} an enlarged view 
of the immediate shock vicinity, around $x\,\simeq\,0.475\,\mathcal{R}_0$ in Fig.~\ref{fig:steepening}(a).  We identify a (double) soliton-like structure \citep{Alsop_1988}, where the particle distribution 
forms a semicoherent cold ring in momentum space (see c.3 in the bottom panel of Fig. \ref{fig:shock}).   A similar structure 
has been reported earlier for infinite planar s hocks~\cite{plotnikov2019}.
The precursor wave is dominated by a linearly polarized X-mode that propagates superluminaly against the upstream plasma.

The spectrum of the precursor wave (Fig \ref{fig:spectrum}), as measured in the Lab frame, exhibits several sharp harmonic peaks around $\omega \simeq 10 \,\omega_{\rm p}'$,
where $\omega'_{\rm p}$ denotes the proper plasma frequency,  measured in the local rest frame of the fluid
in the immediate upstream of the double soliton structure.  These peaks correspond to the lowest harmonics of
the resonant cavity defined by the double soliton structure seen in Fig. \ref{fig:shock}. In good agreement with \cite{plotnikov2019}, the wavelength of the highest peak and the width of the cavity are proportional, $\lambda_{\rm peak}\sim L_{\rm sh}/3$, with a weak dependence of $L_{\rm sh}$ on the upstream magnetization, $\sigma_{\rm u}$.  This implies that the peak frequency, as measured in the Lab frame, should scale as 
the shock Lorentz factor with respect to the Lab frame, $\gamma_{\rm sh|Lab}\sim \sqrt{\sigma_{\rm u}}$  (see appendix).
From our simulations we estimate $\omega_{\rm peak}\approx 1.6\sqrt{\sigma_{\rm u}}\omega'_{\rm p}$ for the highest peak.
Above the peak, the spectrum extends 
up to about $70\,\omega'_{\rm p}$. Below the peak, it cuts off at a frequency below which the wave is trapped by the shock (that is, 
the group velocity is smaller than 
the shock velocity).  The latter can be estimated most easily in the rest frame of the upstream plasma.  In this frame, the dispersion relation 
can be written as \cite{hoshino1991},
\begin{equation}\label{eq:dispersionR}
    \frac{k^{'2}}{\omega^{'2}}= 1-\frac{\omega^{'2}_{\rm p}}{\omega^{'2} - \omega^{'2}_{\rm c}},
\end{equation}
where $\omega'_{\rm c} = \sigma_{\rm u} \omega'_{\rm p}$ is the cyclotron frequency.  
Upon transforming to the Lab frame, the dispersion relation reduces to $k^2 = \omega^2 -\omega_{\rm p}^{'2}$
in the limit $\gamma^2_{\rm sh|u} \gg \sigma_{\rm u}\gg1$ (see appendix), here $\gamma_{\rm sh|u}$ is the shock Lorentz factor measured in the upstream frame. 
The cutoff frequency and k-vector are obtained by equating the group velocity, $v_{\rm g} = d\omega/dk$, with the shock 
velocity with respect to the Lab frame, $v_{\rm sh|Lab}$:
\begin{align}\label{eq:dispersion_Lab}
   \omega_{\rm cut} = \omega_{\rm p}'  \gamma_{\rm sh|Lab} \,  , \qquad k_{\rm cut} =  \omega_{\rm p}'  u_{\rm sh|Lab}.
\end{align}
From the analysis of the data presented in Fig \ref{fig:shock} we estimate $\sigma_{\rm u}=25$ and $\sigma_{\rm u}/\gamma^2_{\rm sh|u}= 6\times10^{-3}$,
from which we obtain $u_{\rm sh|Lab}= 4.3$ upon transforming to the Lab frame.  
The cutoff frequency and k-vector given in Eq. \eqref{eq:dispersion_Lab} are in good agreement with those 
seen in Fig \ref{fig:spectrum}.

Regarding the efficiency of the precursor wave, we find that the X-mode carries a fraction of about $\epsilon_X=10^{-3}$ of the total energy 
dissipated in the shock (the bottom half wave; see Fig. \ref{fig:spectrum}). As found in~\cite{plotnikov2019}, the fraction of incoming kinetic energy converted into precursor waves in the post-shock frame is independent of the magnetization in the limit of $\sigma\gg1$.
The increase in efficiency by a factor of unity ($\lesssim 3$) is consistent with the increase in downstream frame incoming particle kinetic energy across the simulation.   
We thus anticipate $\epsilon_{\rm X}$ to be weakly sensitive to the conditions prevailing in the medium in which the FMS wave propagates. 

%%%%%%%%%%%%%%
% Discussion
%%%%%%%%%%%%%%
We now consider a wave with luminosity $L = c E^2 r^2 /2 =10^{43} L_{43}$ erg/s \footnote{Magnetar bursts span a wide range of luminosities, from $10^{36}$ to $10^{43}$ erg $s^{-1}$, while giant flares, though rare, can reach luminosities as high as $10^{47}$ erg $s^{-1}$ \citep{kaspi2017}. For comparison, we use the normalization 
adopted in \citep{beloborodov23} for the wave power.},
generated near the surface of a magnetar, and assume for simplicity that the wave propagates in the equatorial 
plane of the dipole background field, where $B_{\rm bg}(r) = 10^{15} B_{15} (R/r)^3$ G; $R=10^6$ cm being the stellar radius. 
The background density depends on the pair multiplicity $\mathcal {M}$ in the magnetosphere, which is uncertain but expected 
to be large \cite{beloborodov2021}; we henceforth adopt ${\cal M} = 10^6 {\cal M}_6$ \footnote{We measure the multiplicity with respect to 
the Goldreich–Julian density, rather than the particle number associated with the current flowing in the magnetosphere, $I_0/e\Omega 
\approx \mu \Omega/ec$, as common in the pulsar literature. Here $\mu$ is the dipole moment
of the magnetar and $\Omega$ its rotation frequency.}.
With this normalization, the background number density, $n_{\rm bg} =\rho_{\rm bg}/m_{\rm e}$, can be expressed as 
$n_{\rm bg} = \mathcal {M} n_{\rm GJ} \approx 10^{19} \mathcal {M}_6 B_{15} \Omega (R/r)^3$ cm$^{-3}$, 
where $\Omega$ is the angular velocity of the star, measured
in rad/s, and $n_{\rm GJ} = \Omega B_{\rm bg}/2\pi e c$ is the Goldreich–Julian density. The corresponding 
 plasma frequency of the background pair plasma is $\omega_{\rm p}\,=\,10^{14} \mathcal{M}_6^{1/2} B_{15}^{1/2} (R/r)^{3/2}$ Hz.

The wave propagates nearly undisturbed in the magnetosphere until reaching a radius $r_{\rm s}$ at which the wave breaks. 
This happens when $|E| = B_{\rm bg}/2 $, or $r_{\rm s} \simeq 2\times10^8 B_{15}^{1/2}L_{43}^{-1/4}$ cm.  As discussed above, a shock 
forms and continues evolving with the wave.  It can be shown \cite{beloborodov23} that the Lorentz factor just upstream of 
the shock quickly reaches a maximum value, $\gamma_{\rm u,max}\sim c\sigma_{\rm bg}/\omega r_{\rm s}$ and then gradually declines, roughly as $(r/r_{\rm s})^{-4}$
up to $\sim 3r_{\rm s}$, where the lower half of the wave is completely erased.  The majority of the dissipated energy will be released in the
$x$-ray and $\gamma$-ray bands, and a fraction of $\sim 10^{-3}$ in the form of a precursor wave.
For our choice of parameters, and wave frequency $\nu = \omega/2\pi = 10^4$ Hz, $\gamma_{\rm u,max} \sim 6\times 10^5 (B_{15}\Omega \mathcal{M}_6)^{-1}L_{43}$.  The upstream magnetization
reaches a minimum, $\sigma_{\rm u} = \sigma_{\rm bg}/2\gamma_{\rm u,max} \sim \omega r_{\rm s}/2c \sim 300 B_{15}^{1/2}L_{43}^{-1/4}$, and only slightly increases thereafter \cite{beloborodov23}.  The proper plasma frequency in the shock upstream satisfies $\omega_{\rm p}' = \omega_{\rm p}/\gamma_{\rm u}^{1/2}\approx 10^{8} (r/r_{\rm s})^{1/2} \mathcal{M}_6 B_{15}^{1/4}L_{43}^{-1/8}\Omega^{1/2}$ Hz, and barely changes in the wave dissipation zone, $r_{\rm s} < r < 3r_{\rm s}$.  Adopting the scaling 
we find from the simulations, $\omega_{\rm peak}\approx \sqrt{\sigma_{\rm u}} \, \omega_{\rm p}'$, we anticipate 
the observed spectrum of the precursor wave to appear in the GHz band, as seen in many fast radio bursts, assuming $\mathcal{M}_6\sim 1$.

Whether the precursor wave can escape the inner magnetosphere is unclear at present.   It has been argued that GHz
waves will be strongly damped in the inner magnetosphere by nonlinear decay into Alfv\'en waves \cite{golbarkin2023}, 
steepening or kinetic effects \cite{beloborodov23b,subacchi2024} (but cf. \cite{lyutikov2024}).  
We note, however, that for faint FRBs, like the one emitted by the Galactic magnetar SGR 1935+2154, the 
strength parameter of the precursor wave, $a_0 = e E/2\pi \nu m_{\rm e} c$, is of order a few in the dissipation zone, where the precursor wave is formed,
so the wave is likely to escape \citep{subacchi2024}.
Moreover, the peak frequency of the precursor wave is smaller than the background frequency by a factor $\sqrt{\gamma_{\rm u}/\sigma_{\rm u}} \sim$ a few, so
the wave transitions from the MHD to the kinetic regime, and the damping mechanism considered in \citep{golbarkin2023} needs to be reassessed. 
More generally, the precursor wave is trapped by the kHz FMS wave, and the effect this has on the damping of high power precursor waves 
needs to be studied. To that end, we plan to perform simulations that follow the evolution of the system long after the dissipation of the 
KHz wave is completed.   Finally, it has been shown \cite{beloborodov23} that under magnetar conditions, 
the cooling rate of the upstream plasma entering the shock is comparable to or even shorter than the Larmor frequency.
How this might affect the efficiency of the precursor wave has yet to be determined. 

In summary, we have demonstrated the self-consistent steepening, monster shock formation, and precursor wave emission emerging from a fast magnetosonic wave propagating along a declining background magnetic field. The analytical properties of wave steepening are found to be in good agreement with \emph{ab initio} fully kinetic simulations. The ensuing shock formation leads to efficient dissipation of the bottom part of the wave over the dynamical time of the wave crossing. A fraction of about $10^{-3}$ of the total dissipated energy is imparted to X-modes. The associated spectrum propagating upstream shows pronounced peaks corresponding to harmonics of the cavity forming between two leading solitons at the shock front. Finally, our results open promising avenues to study the fate of electromagnetic pulses propagating in strongly magnetized environments to address self-consistently their dissipation and, if any, the escaping signal. 

\begin{acknowledgments}
\emph{Acknowledgments} 
We thank the anonymous referees for their positive criticism and helpful comments.  We also thank
Andrei Beloborodov, Alex Chen, Hayk Hakobyan, Sasha Philippov, Illya Plotnikov, Lorenzo Sironi and Chris Thompson for insightful discussions. This work was supported by a grant 
from the Simons Foundation (MPS-EECS-00001470-04) to AL. This work was facilitated by the 
Multimessenger Plasma Physics Center (MPPC, NSF grant PHY-2206607). The presented
numerical simulations were conducted on the STELLAR
cluster (Princeton Research Computing). This work was also granted access to the HPC resources of TGCC/CCRT under
the allocation 2024-A0160415130 made by GENCI. 
\end{acknowledgments}

\appendix

\section{Appendix}

\subsection{\label{SM:FMS}Exact solution for a 1D fast magnetosonic wave}

Consider a FMS wave propagating in the positive $x$ direction is a scalar potential background magnetic field of the form 
$\pmb{B}_{\rm bg}(x,z)\,=\,-\nabla \Psi\,=\,-\partial_x\Psi \hat{x} -\partial_z\Psi \hat{z}$, such that in the $x,y$ plane $\partial_x\Psi (x,0)\,=\,0$, so that $\pmb{B}_{\rm bg}(x,0)\,=\,B_{\rm bg}(x) \hat{z}$.  Since $\nabla\times\pmb{B}_{\rm bg}\,=\,0$, $\partial_z B_x (x,0)\,=\,-\partial_z\partial_x \Psi(x,0)\,=\,\partial_x B_{\rm bg}(x)$.
The wave fields are denoted by $\pmb{B}_{\rm w} = B_{\rm w} \hat{z}$, $\pmb{E} = E \hat{y}$.
The equations governing the dynamics of the wave are derived using the general form of the stress-energy tensor:
\begin{equation}
T^{\mu\nu}=(w+b^2)u^\mu u^\nu + (p+b^2/2)g^{\mu\nu} -b^\mu b^\nu,
\end{equation}
where $p$ is the pressure, $w$ the specific enthalpy, and $\sqrt{4\pi} b_\mu= F^\star_{\nu\mu} u^\nu$, here $F^\star$ 
denotes the dual electromagnetic tensor.  In terms of $b^\mu$ and $u^\mu$ it can be expressed as:
\begin{equation}\label{eq:F*}
 F^{\star}_{\mu\nu} =\sqrt{4\pi}( b_\mu u_\nu-b_\nu u_\mu).
\end{equation}
The homogeneous Maxwell's equations read in this notation:
\begin{equation}
\partial_\mu(b^\mu u^\nu-b^\nu u^\mu)=0,\label{induction}
\end{equation}
subject to $b_\mu u^\mu=0$.
For the FMS wave under consideration  $u^\mu=(\gamma, u, 0, 0)$, and $\sqrt{4\pi} b^\mu=(B_x u, B_x\gamma,0,B')$, 
where $B'$ is the $z$ component of the total magnetic field, as measured in the fluid rest frame. In the Lab frame $B=\gamma B^\prime = B_{\rm w} + B_{\rm bg}$.  
From Eq. \eqref{eq:F*} we find 
\begin{equation}\label{eq:E=vB}
E = F_{31}^\star = B' u\quad \Rightarrow  \quad   v= u/\gamma  = E/B,
\end{equation}
here $E = F_{31}^\star$ is the wave electric field.

Denote for short $\tilde{p}=p+B^{\prime 2}/8\pi$ and $\tilde{h}=h+B^{\prime 2}/(4\pi \rho)$, where $h=w/\rho$ is the enthalpy per particle,
the $z$ component of Equation (\ref{induction}) combined with the continuity equation, $\partial_\mu(\rho u^\mu)=0$ (assuming no particle creation), yields: $u^\mu\partial_\mu (B' /\rho)=0$.
Thus, $\xi=B^{'2}/8\pi \rho^2$ is conserved along streamlines.  We choose an equation of state of the form $p=A \rho^{\Gamma}$ for adiabatic index $\Gamma$.
Then 
\begin{eqnarray}
\tilde{p}= A\rho^{\Gamma} + \xi \rho^2, \qquad \tilde{h}= 1 + \frac{\Gamma}{\Gamma-1}A \rho^{\Gamma-1}+ 2\xi \rho.
\end{eqnarray}
Using the equation $\partial_\mu T^{\mu x}=0$, the relation $\partial_z B_x = \partial_x B_{\rm bg}$, and the continuity equation, one finds:
\begin{eqnarray}
\rho \gamma \tilde{h}\frac{\dd u}{\dd t}+\rho \gamma u \frac{\dd\tilde{h}}{\dd t}+\partial_x \tilde{p}= \frac{B\partial_x B_{\rm bg} }{4\pi},  \label{eq:E-P-Eq}\\
\frac{\dd}{\dd t}(\rho\gamma)+\rho\gamma\partial_x v=0, \label{eq:cont}
\end{eqnarray}
here $\dd/\dd t=\partial_t + v \partial_x$ is the convective derivative, and $v=u/\gamma$.  
The fast magnetosonic speed $a$ is defined by
\begin{equation}\label{eq:a}
a^2=\frac{\partial \ln \tilde{h}}{\partial \ln \rho}= \frac{\Gamma p + (B^{'2}/4\pi )}{\rho\tilde{h}}.
\end{equation}
For a cold plasma, $p=0$, it reduces to
\begin{equation}\label{eq:a_cold}
a^2 = \frac{\sigma}{1+\sigma},
\end{equation}
where
\begin{equation}
\sigma = \frac{B^{'2}}{4\pi \rho} = 2\xi \rho,
\end{equation}
is the magnetization of the fluid.   
In terms of the variable $\dd\chi= a \dd\ln \rho$,  Eqs. \eqref{eq:E-P-Eq} and \eqref{eq:cont}  can be rewritten 
in the form 
\begin{eqnarray}
a(v\partial_t \chi+\partial_x\chi)+\gamma^2(\partial_t v+v\partial_x v) &=& \frac{1}{4\pi}B\partial_x B_{\rm bg},\\
(\partial_t \chi+v \partial_x\chi)+a\gamma^2( v \partial_t v+\partial_x v) &=& 0,
\end{eqnarray}
where the relations  $\dd\tilde{h}=a\tilde{h} \dd\chi$, $\dd\tilde{p}= \rho \dd\tilde{h} = a \rho\tilde{h}\dd\chi$, and $\dd u=\gamma^3 \dd v$ have been employed.
Further manipulation of the latter equations yields 
\begin{equation}
\partial_t J_\pm + v_{\pm} \partial_x J_\pm= \frac{\pm B\partial_x B_{\rm bg}}{4\pi \rho\gamma^2 \tilde{h}(1\pm av)},
\label{dJ_pm}
\end{equation}
here 
\begin{equation}
J_\pm=\chi\pm \lambda,\quad \lambda\equiv \frac{1}{2 }\ln\left(\frac{1+v}{1-v}\right),
\end{equation}
are the Riemann invariants of the system, and
\begin{equation}
v_{\pm}=\frac{dx_\pm}{dt}=\frac{v\pm a}{1\pm a v}
\end{equation}
are the velocities of the characteristics $C_\pm$ on which $J_\pm$ propagate, as measured in the Lab frame.

Consider now a wave injected at time $t=0$ from a source located at $x=x_0$ and moving rightward. 
Since ahead of the wave $v=0$, and the background field is independent of time, $J_- $ satisfies 
\begin{equation}
- a_{\rm bg} \partial_x J_- = - \partial_x \ln B_{\rm bg},
\end{equation}
where $a_{\rm bg}^2\,=\, \sigma_{\rm bg}/(\sigma_{\rm bg}+1)$.  To order $O(\sigma_{\rm bg}^{-1})$  the solution is $J_- = \ln (B_{\rm bg}/B_{0}) \equiv \chi_{\rm bg}$, here $B_{0} = B_{\rm bg}(x_0)$
is the background magnetic field at the source.  
Then, $J_+ = 2\lambda + \chi_{\rm bg}$, and   Eq. (\ref{dJ_pm}) for $J_+$ reduces to
\begin{equation}
\partial_t \lambda+ v_{+} \partial_x \lambda= \frac{BB_{\rm bg}/(4\pi \rho\gamma^2 \tilde{h}) -a-v}{2(1+av)} \partial_x\ln B_{\rm bg}.
\label{eq:dlambda}
\end{equation}

To find $a(\lambda)$, we use the relations  $a^2=2\xi \rho /(1 +2\xi \rho)$ (Eq. \ref{eq:a}), suitable for cold plasma, 
and ${\rm d}\chi=a\,{\rm d}\ln \rho$. Integrating from $\chi_{\rm bg}$ to $\chi$, one obtains
\begin{eqnarray}
a(\lambda) &=& \tanh (\lambda/2+c),\\
v_+(\lambda) &=& \tanh (3\lambda/2+c),
\end{eqnarray}
where the constant $c$ is determined from the condition $a(\lambda = 0) = a_{\rm bg}$: $c= \ln (\sqrt{\sigma_{\rm bg}} +\sqrt{\sigma_{\rm bg}+ 1})$.  
Note that $\lambda=0$ implies $v=0$ and $E=0$, hence,
the magnetic field equals the background field.  
The magnetization can be readily computed using Eq. \eqref{eq:a_cold}: $\sigma = a^2/(1-a^2) = \sinh^2(\lambda/2 + c)$.
From this, one finds:
 \begin{equation}\label{eq:B'}
 \frac{B'}{B_{\rm bg}} = \frac{\rho}{\rho_{\rm bg}} =\frac{\sigma}{\sigma_{\rm bg}}= \frac{\sinh^2(\lambda/2+c)}{\sinh^2(c)} \xrightarrow[\lambda+c \, \gg 1]{} \sqrt{\frac{1+v}{1-v}}.
\end{equation}
With the above results, the right-hand side of Eq. \eqref{eq:dlambda} can be expressed as a function of $\lambda$. Let us denote for short
$A(\lambda) = \tfrac{1}{2}(B B_{\rm bg}/4\pi \rho\gamma^2 \tilde{h} -a-v)/(1+av)$.  The evolution of $\lambda$ along the characteristic $C_+$ 
is governed by the coupled ODEs
\begin{eqnarray}
    \frac{\dd\lambda}{\dd t} \Big|_{C_+} &=& A(\lambda) \partial_x \ln B_{\rm bg}, \label{eq:d_lambda}\\
     \frac{\dd x}{\dd t} \Big|_{C_+} &=& v_+(\lambda) = \tanh(3\lambda/2 +c), \label{eq:d_x_+}
\end{eqnarray}
subject to the boundary conditions $x_+(t=0) = x_0$, $\lambda(t_{\rm i},x_0) = \lambda_{\rm i}$, here $\lambda(t,x_0)\equiv \lambda_0(t)$ is a specified injection function.
In case of a uniform field, $\partial_x \ln B_{\rm bg}=0$, $\lambda$ is conserved along $C_+$ and the solution can be expressed as 
$\lambda(x,t) = \lambda[\xi(x,t)]$, with $\xi(x,t)$ given implicitely by
\begin{equation}\label{eq:xi}
 \xi=t-\frac{x}{\tanh[3\lambda_0(\xi)/2 + c]}.
\end{equation}

Realistic magnetospheres are nearly force-free, that is, $\sigma_{\rm bg} \gg 1$.  In this limit, we can expand $A(\lambda)$ and $v_+(\lambda)$ to 
obtain a more convenient set of equations.  In terms of $\kappa = e^\lambda$ we obtain: 
\begin{equation}\label{eq:A,v_+}
  A(\kappa) = \frac{2 \kappa^2(\kappa^2-1) \sigma_{\rm bg}}{4\kappa^3\sigma_{\rm bg} +1}, \quad v_+(\kappa) = \frac{4\kappa^3\sigma_{\rm bg} -1}{4\kappa^3\sigma_{\rm bg} +1}.
\end{equation}
We consider background magnetic field profiles of the form $B_{\rm bg}(x)\,=\,[1+(x-x_0)/\mathcal{R}_0]^{-p}$, with $p=1$ and $x_0=0$ chosen 
in the simulations presented in the main part.
For convenience and without loss of generality, we choose the origin to be 
located at $x_0\,=\,\mathcal{R}_0$, and normalize time and length to $x_0$; $x \to x/x_0$, $t\to t/x_0$. 
Equations \eqref{eq:d_lambda} and \eqref{eq:d_x_+} then take the form:
\begin{eqnarray}
    \frac{\dd\kappa}{\dd t}\Big|_{C_+} &= & \frac{p}{x} \, \frac{2 \kappa^2(\kappa^2-1) \sigma_{\rm bg}}{4\kappa^3\sigma_{\rm bg} +1}, \label{eq:dk/dt} \\
     \frac{\dd x}{\dd\kappa}\Big|_{C_+} &=& %\frac{dx_+}{dt}\frac{dt}{d\kappa} =   
     \frac{x(4\kappa^3\sigma_{\rm bg} -1)}{2p\sigma_{\rm bg}\kappa^2(\kappa^2 - 1)}.
\end{eqnarray}
To find $\sigma_{\rm bg}$, note that since $B_{\rm bg}/\rho_{\rm bg}=$ const, $\dd\ln \sigma_{\rm bg}/\dd x\,=\,\dd\ln B_{\rm bg}/\dd x = -p/x$, from 
which one obtains:
\begin{equation}
 \frac{\dd\sigma_{\rm bg}}{\dd\kappa} = \frac{\dd\sigma_{\rm bg}}{\dd x}\frac{\dd x}{\dd\kappa} = \frac{4\kappa^3 \sigma_{\rm bg} -1}{2\kappa^2(1-\kappa^2)}.
\end{equation}
The solution is
\begin{equation}\label{eq:sig_bg}
 (1-\kappa^2) \sigma_{\rm bg} = \frac{ 1}{2\kappa} + (1-\kappa_{\rm i}^2)\sigma_0 - \frac{1}{2\kappa_{\rm i}} \equiv \frac{1}{2\kappa}+\alpha_{\rm i},
\end{equation}
where $\kappa_{\rm i} = \kappa(t_{\rm i})$, and $\sigma_0 = \sigma_{\rm bg}(k_{\rm i})$.  The magnetization of the plasma is given by 
$\sigma (\kappa) = \kappa \sigma_{\rm bg}(\kappa) = (1/2 + k\alpha_{\rm i})/(1-\kappa^2)$ (Eq. \ref{eq:B'}). It declines monotonically with $\kappa$
as the wave propagates.
The solution for $x$ is obtained now using $\sigma_{\rm bg}(x) = \sigma_0x^{-p}$:
\begin{equation}\label{eq:x_p}
    x^p = \frac{2\kappa(1-\kappa^2)\sigma_0}{1+2\kappa \alpha_{\rm i}}.
\end{equation}
Substituting $\sigma_{\rm bg}(\kappa)$ and $x(\kappa)$ into Eq. \eqref{eq:dk/dt} yields
\begin{equation}
    \frac{\dd t}{\dd\kappa} = - \frac{1}{p} \left[ \frac{2\kappa}{1-\kappa^2} +\frac{1}{\kappa (1+2\kappa \alpha_{\rm i})} 
    \right]\left[\frac{2\kappa(1-\kappa^2)\sigma_0}{1+2\kappa\alpha_{\rm i}} \right]^{1/p}.
\end{equation}
The solution can be expressed as
\begin{equation}\label{eq:kap(t)}
\begin{split}
  & t(\kappa) -t_{\rm i} = -\frac{(2\sigma_0)^{1/p}}{p} \times\\
  & \int^k_{k_{\rm i}} \dd\kappa' \left[ \frac{2\kappa'}{1-\kappa'^2} +\frac{1}{\kappa' (1+2\kappa' \alpha_{\rm i})} 
    \right]\left[\frac{2\kappa'(1-\kappa'^2)}{1+2\kappa'\alpha_{\rm i}} \right]^{1/p}
   \end{split}
\end{equation}

This solution can be inverted to yield $\kappa(t)$. Substituting into Eq.\eqref{eq:x_p} yields the trajectory of the characteristic, $x(t)$, that satisfies 
$x(t_{\rm i})=1$.  The family of all characteristics is labelled by the injection time $t_{\rm i}$. It can be expressed as $x(t,t_{\rm i})$, with $x(t_{\rm i},t_{\rm i})=1$.
The injection function gives the corresponding values of $\kappa$ at $t_{\rm i}$.  For injection of  a periodic wave 
at the boundary ($x=1$), with electric field $E(1,t_{\rm i}) = a B_{0} \sin(\omega t_{\rm i})$,  Eqs. \eqref{eq:E=vB} and \eqref{eq:B'} yield:
\begin{equation}
    \kappa_{\rm i}(t_{\rm i}) = \sqrt{1+2a\sin(\omega t_{\rm i})}.
\end{equation}

Crossing of $C_+$ characteristics %is expected when $dv_+(t_{\rm i},x_0)/dt_{\rm i} > 0$.  It 
is determined from the condition $\partial x(t,t_{\rm i})/\partial t_{\rm i}=0$,
where the partial derivative is taken at a constant time $t$.  However, this equation admits multiple solutions.  A shock will initially form at the 
first crossing, that is, at the earliest time, defined by $\partial t(\kappa,t_{\rm i}) /\partial t_{\rm i}=0$. The solution of the two equations above yields 
a unique value, $\kappa_{\rm s}$, from which the location and time of shock formation can be computed.  The plasma 4-velocity at this location is given
by $u(\kappa_{\rm s}) = (\kappa_{\rm s}^2-1)/2\kappa_{\rm s} \approx -1/2\kappa_{\rm s}$.

\begin{figure}
\centering
    \includegraphics[width=1.\columnwidth]{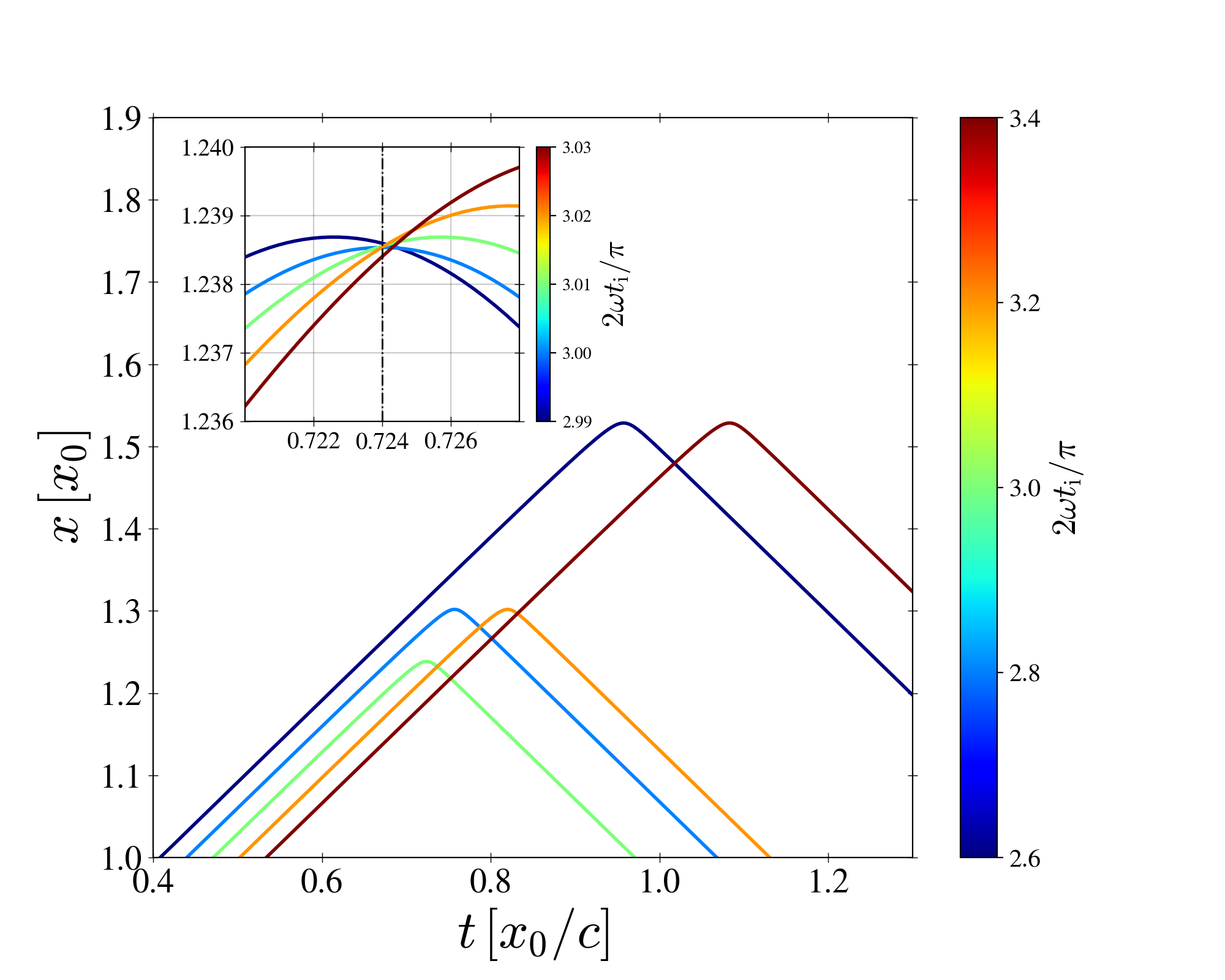}
    \caption{A sample of wave characteristics with initial values at the wave trough, $2\omega t_{\rm i}/\pi = 2.6, 2.8, 3, 3.2, 3.4$, and
    $\sigma_0=1600$, $a=0.4$, $\omega=10$, $p=1$. The inset shows a zoom in around the time of shock birth (shown by the 
    vertical dashed line) for more closely spaced characteristics in the interval %$2.97 \le 2\omega t_{\rm i}/\pi \le 3.02$ 
    $2.99 \le 2\omega t_{\rm i}/\pi \le 3.03$}
 \label{fig:charc}
\end{figure}

For illustration, consider $p=1$. For this choice Eq. \eqref{eq:kap(t)} can be readily integrated, yielding
\begin{equation}\label{eq:t_kappa_general}
   t(\kappa) -t_{\rm i} = 2\sigma_0 [G(\kappa)-G(\kappa_{\rm i})],
\end{equation}
where 
\begin{equation}
\begin{split}
    G(\kappa) &= \frac{1}{2\alpha_{\rm i}(1+2\alpha_{\rm i} \kappa)} + \frac{2+3\alpha_{\rm i}\kappa -2\alpha_{\rm i}^2\kappa^2}{4\alpha_{\rm i}^3}\\
    & - \frac{1}{8\alpha_{\rm i}^3}\left[4\ln(1+2\alpha_{\rm i}\kappa) + \frac{1}{1+2\alpha_{\rm i}\kappa}\right].
\end{split}
\end{equation}
To estimate the location of the caustic (if exists) for a characteristic with initial value $\kappa_{\rm i}$ we use Eq. \eqref{eq:x_p} to obtain 
$\partial x/\partial t_{\rm i} = (\partial x/\partial \kappa_{\rm i})(\partial \kappa_{\rm i}/\partial t_{\rm i})=0$.  Assuming $\partial \kappa_{\rm i}/\partial t_{\rm i} \ne 0$,
one finds: 
\begin{equation}\label{eq:dx=0}
 \frac{\partial \kappa_{\rm c}}{\partial\kappa_{\rm i} }= \frac{4\alpha_{\rm i}\kappa_{\rm c}^2\kappa_{\rm i}}{(4\alpha_{\rm i}\kappa_{\rm c}^3-1) } ,
\end{equation}
here $\kappa_{\rm c}(\kappa_{\rm i})$ denotes  the value of $\kappa$ at the caustic for the characteristic $\kappa_{\rm i}$.
Now, for a typical case we anticipate $\alpha_{\rm i}\gg 1$ and $\kappa_{\rm c} \ll 1$. In addition, since the caustic occurs 
 near the turnaround point, $v_+=0$, at which $4\kappa^3\sigma_{\rm bg} =1 \approx 4\kappa^3 \alpha_{\rm i}$ (see Eqs. \ref{eq:A,v_+} and \ref{eq:sig_bg}),
 we anticipate $4\kappa_{\rm c}^3\alpha_{\rm i} \approx 1$.
 Expanding in powers of $\alpha_{\rm i}^{-1}$  and $\kappa$, we find 
 \begin{equation}\label{eq:t_k_2}
   t(\kappa_{\rm c})-t_{\rm i} \approx \frac{1}{1-\kappa_{\rm i}^2}\left[  \kappa_{\rm i}^2 - \kappa_{\rm c}^2 +\frac{1}{2\alpha_{\rm i} \kappa_{\rm c}} \right ]. %-\frac{1+6\kappa_{\rm i}^2}{2\alpha_{\rm i}\kappa_{\rm i}}\right]
\end{equation}
The shock first appears at the earliest crossing, that is, at $\kappa_{c} = \kappa_{\rm s}$ for which $t(\kappa_{\rm s})$ is a minimum.
Differentiating Eq. \eqref{eq:t_k_2} at a constant $t$, and substituting $\partial\kappa_{\rm c}/\partial\kappa_{\rm i}$ from Eq. \eqref{eq:dx=0}, gives 
\begin{equation}\label{eq:k_sh_2}
\begin{split}
  4\alpha_{\rm i}\kappa^3_{\rm s}(t_{\rm i}) & =
 \frac{(2-\kappa_{\rm i}^2)\frac{\partial\kappa^2_{\rm i}}{\partial t_{\rm i}} +(1-\kappa_{\rm i}^2)^2}
  {\kappa_{\rm i}^2\frac{\partial\kappa^2_{\rm i}}{\partial t_{\rm i}} +(1-\kappa_{\rm i}^2)^2}  
     \approx 1 - 2\omega \cot(\omega t_{\rm i}),
  \end{split}
\end{equation}
using $\kappa^2_{\rm i}(t_{\rm i}) = 1+2a\sin(\omega t_{\rm i}) $, $ \partial\kappa^2_{\rm i}/\partial t_{\rm i} = 2\omega a \cos(\omega t_{\rm i})$. 
Comparing Eq. \eqref{eq:k_sh_2} with Eq. \eqref{eq:dx=0}, we obtain the compression factor,
\begin{equation}\label{eq:est_k_sh}
    \kappa_{\rm s} = \left( \frac{\omega^2}{48 a^3 \sigma_0^2}\right)^{1/4},
\end{equation}
and the  time of shock birth: 
\begin{equation}\label{eq:est_tsh}
  t_{\rm s}\equiv  t(\kappa_{\rm s})  = \frac{3\pi}{2\omega} +\frac{1}{2a}-1 + \frac{1}{8 a \sigma_0 \kappa_{\rm s}}.
\end{equation}
The corresponding  plasma 4-velocity is $u_{\rm s} = (\kappa_{\rm s}^2-1)/2\kappa_{\rm s}\approx -1/2\kappa_{\rm s} = -(a/\omega^2)^{1/4} \sqrt{\sigma_{\rm bg}} $,
here  $\sigma_{\rm bg}(\kappa_{\rm s}) \approx 2a\sigma_0$ (see Eq. \ref{eq:sig_bg}).  
The magnetization of the flow at this location is $\sigma_{\rm s} =\kappa_{\rm s} \sigma_{\rm bg}(\kappa_{\rm s}) \gg 1$.

\begin{figure}
  \centering
  \includegraphics[width=1\columnwidth]{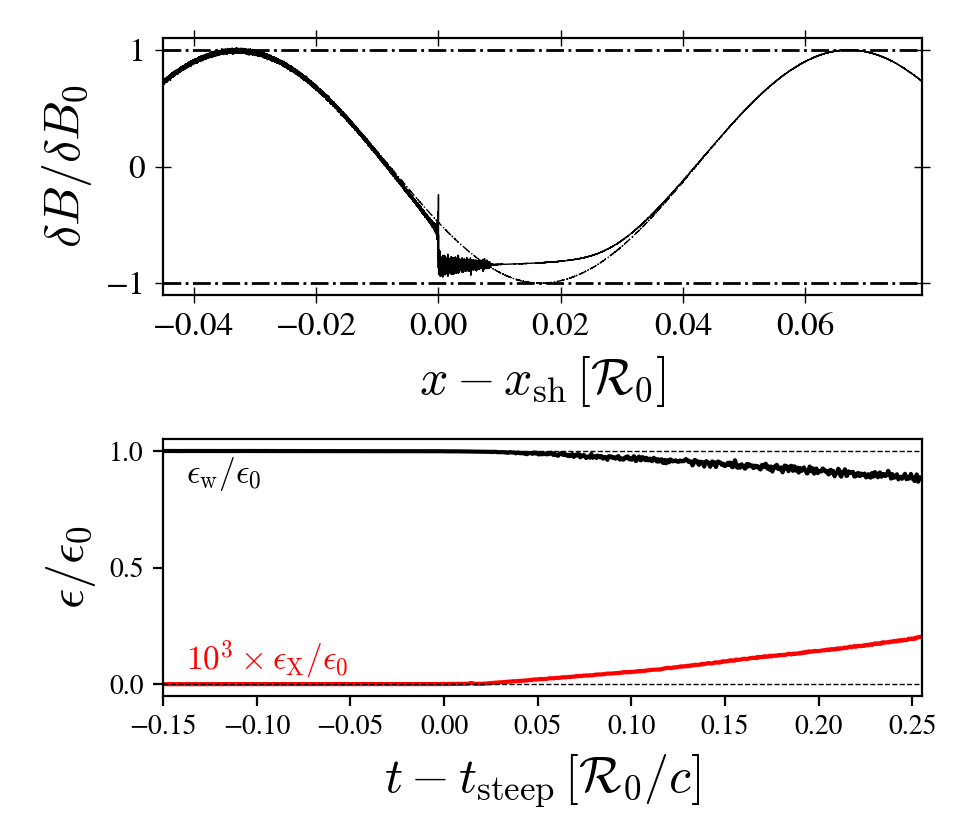}
  \caption{(top) Comparison between the wave profile before and after steepening. The final profile is taken at $t-t_{\rm steep}\,=\, 0.25\,\mathcal{R}_0/c$. (Bottom) Evolution of the normalized wave energy (black) and energy in the X-modes (red). The X-mode profile is multiplied by $10^3$ for readability.}
  \label{fig:dissipation}
\end{figure}

Figure \ref{fig:charc} exhibits sample characteristics computed from Eqs. \eqref{eq:x_p} and \eqref{eq:t_kappa_general}
for $p=1$, $a=0.4$, $\omega =10$, and $\sigma_0=1600$, injected at the indicated times (around the wave trough).  %$2\omega t_{\rm i}/\pi = 2.6, 2.8, 3, 3.2, 3.4$.  
As seen, each characteristic turns around  at $\kappa= (4\sigma_{\rm bg})^{-1/3}$, at which $v_+$ changes sign (Eq. \ref{eq:A,v_+}).
The inset shows a zoom-in around the location of shock birth; it indicates $\kappa_{\rm s}=0.0590$, $t_{\rm s}=0.724$, $x_{\rm s} = 1.238$.
These values are in excellent agreement with the above estimates ($\kappa_{\rm s} =0.0597$ from Eqs. \ref{eq:est_k_sh}, $t_{\rm s} = 0.725$ from Eq. \ref{eq:est_tsh}
and $x_{\rm s} = 1.241$).  In units of the wavelength $\lambda_{\rm w} =2\pi/k$ we have: $x_{\rm s}/\lambda_{\rm w} = x_{\rm s}\omega/2\pi \approx  2$.  The corresponding 
plasma 4-velocity is $u_{\rm s} = -1/2\kappa_{\rm s} \approx -8$.  These numbers are in rough agreement with the simulations, although we stress that
the comparison is not perfect since, in the simulations, we adopt a constant density for numerical convenience rather than invoking $B_{\rm bg}/\rho_{\rm bg} =$ const.

Once the shock forms, it gradually grows in strength as the plasma upstream of the shock continues
accelerating towards the source (since $\kappa$  declines).  To compute the shock velocity, we use the jump conditions 
of a strongly magnetized shock. Since  the magnetization of the upstream plasma is large,
$\sigma_{\rm u}\approx \kappa_{\rm u}\alpha_{i,\rm u} \gg1$, here $\alpha_{i,\rm u} = (1-\kappa_{i,\rm u}^2)\sigma_0$ and $\kappa_{i,\rm u}$ is the initial value 
of the characteristic that crosses the shock from the upstream.  In this limit, the velocity of the downstream flow, as measured 
in the shock frame, is $u'_{\rm d} = -\sqrt{\sigma_{\rm u}}$, where, henceforth, prime denotes quantities measured in the shock frame.  
Mass conservation implies $\rho_{\rm d} u_{\rm d}' =\rho_{\rm u} u'_{\rm u}$. Using Eq. \eqref{eq:B'}, one finds
$u'_{\rm u} = -\kappa_{\rm d}\sqrt{\sigma_{\rm u}}/\kappa_{\rm u}$.  Transforming to the Lab frame we obtain: $u_{\rm u} = \gamma'_{\rm u}\gamma_{\rm sh|Lab}(v'_{\rm u} + v_{\rm sh|Lab})$, where 
$u_{\rm u} = (\kappa_{\rm u}^2-1)/2\kappa_{\rm u}$.  Combining the above results yields the shock velocity in the Lab frame:
\begin{eqnarray}\label{eq:u_sh}
\begin{split}
    u_{\rm sh|Lab} &= \frac{(\kappa_{\rm u}^2+1)\kappa_{\rm d}\sqrt{\sigma_{\rm u}}}{2\kappa_{\rm u}^2} + \frac{(\kappa_{\rm u}^2-1)\sqrt{\kappa^2_{\rm d}\sigma_{\rm u}+ \kappa_{\rm u}^2}}{2\kappa_{\rm u}^2}\\ 
   &\approx \kappa_{\rm d}\sqrt{\sigma_{\rm u}},
    \end{split}
\end{eqnarray}
noting that $\kappa_{\rm d}^2\sigma_{\rm u}/\kappa_{\rm u}^2 = \kappa_{\rm d}^2 \alpha_{\rm i}/\kappa_{\rm u} \gg 1$.
Thus, it is seen that the shock propagates in the positive $x$ direction, that is, opposite to the direction of plasma motion.  
To find the evolution of the wave subsequent to shock formation, one needs to solve the characteristics equations, Eqs. \eqref{eq:d_lambda}-\eqref{eq:d_x_+},
coupled to the equation of the shock trajectory, $\dd x_{\rm s}/\dd t = u_{\rm sh|Lab}/\sqrt{u_{\rm sh|Lab}^2 +1}$. We shall not present such analysis here. 
Instead, the shock structure and propagation are illustrated in Fig.~\ref{fig:dissipation} from the kinetic simulations. As the main text explains, we launch a periodic FMS wave from $x=0$, letting it propagate along $+\hat{x}$ in a cold pair plasma with a declining background magnetic field, constant density, and large initial background magnetization. The weakly declining background magnetic field is set to be static and external for convenience. The top panel shows a comparison between the wave profile after and before saturation. A kinetic scale fluid discontinuity, a shock, is visible at $x-x_{\rm sh} = 0$. The time evolution of the power of the ensuing wave propagating upstream of the shock, dominated by X-modes, is illustrated in the bottom panel.

\begin{figure}
\centering
    \includegraphics[width=1.\columnwidth]{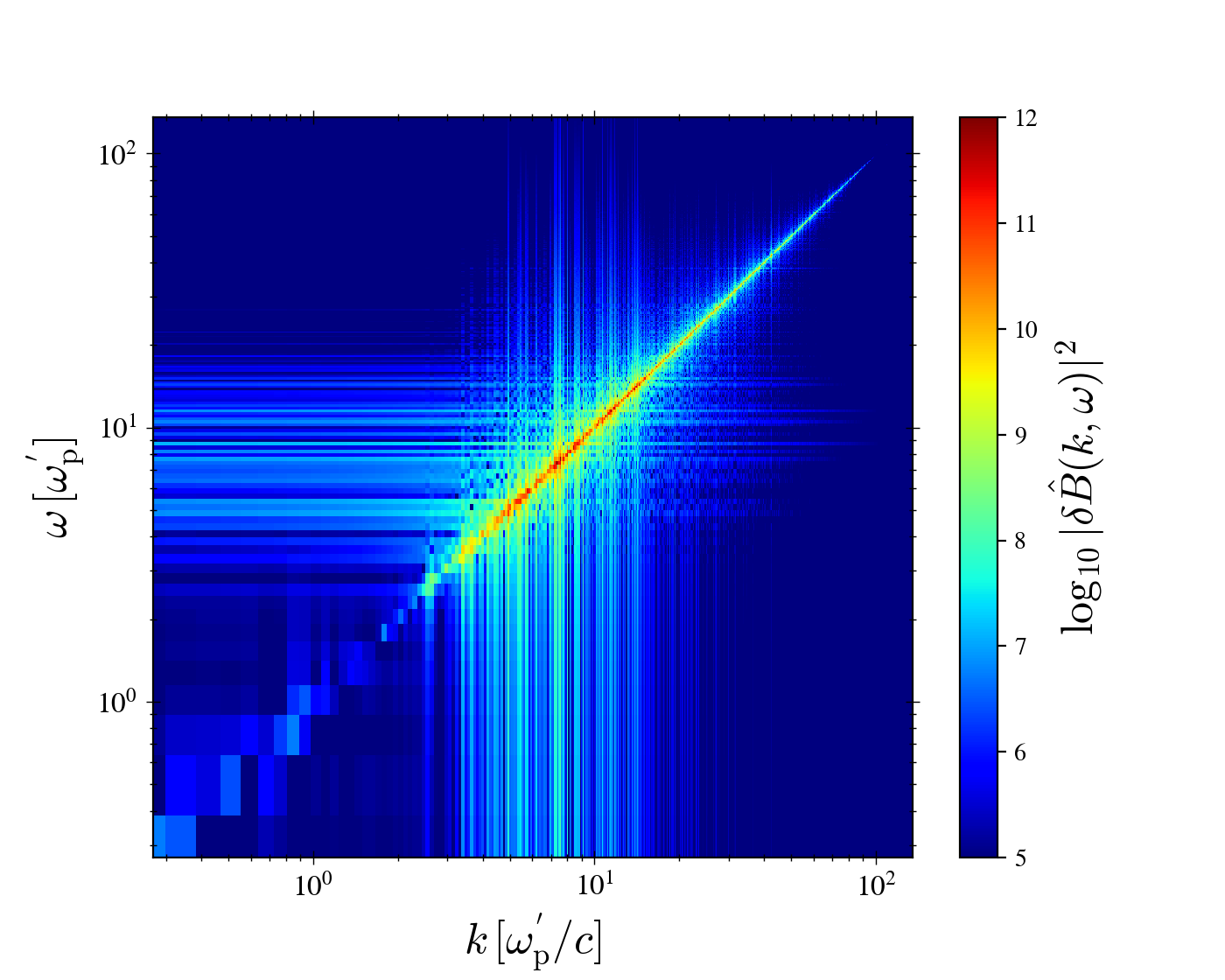}
    \caption{Full $(\omega,k)$-Fourier spectrum of the X-modes at the end of the simulation in units of the upstream plasma frequency. The spectrum is measured in a fixed region in the immediate upstream of the shock and in the laboratory frame.}
 \label{fig:fourier_X_modes}
\end{figure}

\subsection{Dispersion relations}

In the rest frame of the fluid, the dispersion relations of the precursor wave read:
\begin{equation}\label{eq:dispersionR_SM}
    \frac{k^{'2}}{\omega^{'2}}= 1-\frac{\omega^{'2}_{\rm p}}{\omega^{'2} - \omega^{'2}_{\rm c}},
\end{equation}
where $\omega_{\rm p}'$ is the proper plasma frequency, and $\omega'_{\rm c}= \omega'_{\rm p}\sqrt{\sigma}$.
Transforming to the laboratory frame, one obtains,
\begin{eqnarray}
k &=& \gamma_{\rm u} (k' +v_{\rm u} \omega')\,, \\
\omega &=& \gamma_{\rm u} (\omega' + v_{\rm u} k').
\end{eqnarray}

Inverting the transformation and using Eq. \eqref{eq:dispersionR_SM}, yields:
\begin{equation}
  k^2 = \omega^2 -\frac{\omega^{'2}_{\rm p} \gamma_{\rm u}^2(\omega - k v_{\rm u})^2}{\gamma_{\rm u}^2(\omega-k v_{\rm u})^2 -\sigma\omega_{\rm p}^{'2}}.
\end{equation}
Noting that $k v_{\rm u} <0$, we deduce that for $\sigma \ll \gamma_{\rm u}^2$, $\gamma_{\rm u}^2(\omega-k v_{\rm u})^2 \gg \sigma\omega_{\rm p}^{'2}$, so that 
to a good approximation $k^2 = \omega^2 - \omega_{\rm p}^{'2}$.  The group velocity in the Lab frame can be readily computed now:
\begin{equation}\label{eq:SM:v_g}
 v_{\rm g}(\omega) = \frac{\dd\omega}{\dd k} = \sqrt{1-\frac{\omega_{\rm p}^{'2}}{\omega^2}}.
\end{equation}
The wave will be trapped by the shock if $v_{\rm g} \le v_{\rm sh|Lab}$, or
\begin{equation}\label{eq:SM:k_cut}
 \omega\le \omega_{\rm cut}  = \gamma_{\rm sh|Lab} \, \omega'_{\rm p}; \quad k\le k_{\rm cut}= u_{\rm sh|Lab} \, \omega'_{\rm p}\,.
\end{equation}
Figure~\ref{fig:fourier_X_modes} shows the dispersion relation of the X-modes in the immediate shock upstream, obtained 
from the simulations, in units of $\omega^{'}_{\rm p}$. We notice that the branch closely corresponds to light modes, as expected from Eq. \eqref{eq:SM:v_g}, and that the shock 
imposes a low frequency/wavelength cutoff, in agreement
with Eq. \eqref{eq:SM:k_cut}. Note that the dispersion relation is computed in the laboratory frame and not in the upstream frame, as in Eq.~\eqref{eq:dispersionR_SM}.

\bibliography{apssamp}

\end{document}